# INTEGER POLYNOMIAL FACTORIZATION BY RECOMBINATION OF REAL FACTORS: RE-EVALUATING AN OLD TECHNIQUE IN MODERN ERA

SHAHRIAR IRAVANIAN

Abstract. Polynomial factorization over $\mathbb{Z}$ is of great historical and practical importance. Currently, the standard technique is to factor the polynomial over finite fields first and then to lift to integers. Factorization over finite fields can be done in polynomial time using Berlekamp or Cantor-Zassenhaus algorithms. Lifting from the finite field to $\mathbb{Z}$ requires a combinatorial algorithm. The van Hoeij algorithm casts the combinatorial problem as a knapsack-equivalent problem, which is then solved using lattice-reduction (the LLL algorithm) in polynomial time, which is implemented in many computer algebra systems (CAS).

In this paper, we revisit the old idea of starting with factorization over $\mathbb{R}$ instead of a finite field, followed by recombination of the resulting linear and quadratic factors. We transform the problem into an integer subset sum problem, which is then solved using the Horowizt-Sinha algorithm. This algorithm can factor a random integer polynomial of degree $d$ in a time complexity of $O(2^{d/4})$.

While the resulting algorithm is exponential, consistent with the integer subset sum problem being in NP, it has a few advantages. First, it is simple and easy to implement. Second, it is almost embarrassingly parallelizable. We demonstrate this by implementing the algorithm in a Graphic Processing Unit (GPU). The resulting code can factor a degree 100 polynomial is a few tenths of a second, comparable to some standard CAS. This shows that it is possible to use current hardware, especially massively parallel systems like GPU, to the benefit of symbolic algebra.

## 1. Introduction

Factorization of polynomials with integer coefficients is of significant historical and practical importance, with a history going back to Isaac Newton. Further work during the eighteenth and nineteenth centuries resulted in the Schubert-Kronecker algorithm, which remained the mainstay of polynomial factorization until the advent of digital computers when it was realized that the Schubert-Kronecker algorithm did not perform well. This realization started extensive research to find fast factorization algorithms [1, 2].

We can consider the two decades from 1965 to 1985 the golden era of polynomial factorization when most of the current algorithms used in computer algebra systems (CAS) originated. Berlekamp discovered a polynomial-time algorithm for factorization over finite fields, where the coefficients are in $\mathbb{Z}_m$ for $m$ a prime [3]. Berlekamp's discovery showed the importance of working in finite fields and ignited research efforts to extend the results to polynomials with integer coefficients.

Zassenhaus and co-workers successfully solved the problem of finding a factorization over $\mathbb{Z}$ based on factorization over $\mathbb{Z}_m$ [4, 5]. The high-level summary of the roundabout or Berlekamp-Zassenhaus algorithm is as follows:

1. Choose a prime $m$.

2. Factor $p(x)$ over $\mathbb{Z}_m$ using either the Berlekamp algorithm or, more commonly, the Cantor-Zassenhaus distinct degree/equal degree algorithms.
3. Lift factorization from $\mathbb{Z}_m$ to $\mathbb{Z}_{m^2}, \mathbb{Z}_{m^3}, \ldots$ using Hensel lifting.
4. Let $p = p_1 p_2 \ldots p_n \bmod m^\alpha$ for $\alpha \in \mathbb{Z}$. Try $2^n$ test vectors $v \in \{0, 1\}^k$ such that $q = \prod p_i^{v_i}$ divides $p$ in $\mathbb{Z}$. This search process is sometimes called *recombination*.

Step 4 in the roundabout algorithm suffers from exponential worst-case complexity. However, in practice, it has an average polynomial time complexity for most random polynomials and is the base algorithms implemented in various CAS.

In 1982, Lenstra, Lenstra, and Lovasz found a polynomial-time algorithm for integer polynomial factorization and introduced the LLL lattice-reduction algorithm [6]. However, its $O(d^{10})$ complexity, where $d$ is the degree of the input polynomial, makes it unsuitable for most practical applications.

The focus in the 1990s shifted to randomized and symbolic-numeric algorithms. For example, the Shoup-Kaltofen algorithm achieved subquadratic-time in factoring polynomials over finite fields [7].

The next breakthrough was the discovery of the van Hoeij algorithm in the early 2000s [8]. The van Hoeij algorithm replaces step 4 of the roundabout algorithm with a knapsack-equivalent problem, which is then solved using lattice-reduction of the LLL algorithm. The resulting algorithm is polynomial and practical. Nevertheless, it is complicated and difficult to implement.

In contrast to the sophisticated mathematics used in these algorithms based on factorization over finite fields first, there is a simple and essentially trivial algorithm based on finding a subset of the polynomial roots that would result in a factor with integer coefficients. The main idea is to replace steps 1-3 of the roundabout algorithm with factorization of $p(x)$ over $\mathbb{R}$. Since $p(x) \in \mathbb{Z}[x] \subset \mathbb{R}[x]$, such factorization to linear and quadratic factors is always possible. Then, the algorithm follows, similar to step 4, by trying $2^n$ combinations of the factors, where $n$ is the number of the factors. This paper calls this algorithm the Real-Factors-Recombination (RFR) algorithm.

It is almost certain that the algorithm we are presenting, or a similar one, was known to the mathematicians who developed various factorization algorithms. However, it was likely dismissed as impractical due to its exponential time complexity. This dismissal was reasonable in the 1960s and 1970s, when the speed and memory capacity of most computers would have made such an algorithm impractical.

The classic roundabout (Berlekamp-Zassenhaus) and RFR algorithms are exponential in the number of factors used for recombination. For RFR, the factors are linear and quadratic; therefore, $\frac{d}{2} \leq n \leq d$. On the other hand, for the roundabout algorithm, a reasonable hope is that some factors have higher degrees and, therefore, there are fewer of them.

However, as we will see, we can do better than $O(2^n)$ that is naively expected. We also cast the problem as a knapsack-equivalent one, but instead of going the way of the LLL algorithm, as van Hoeij algorithm does, we apply a more straightforward and older method. Using the Horowitz-Sahni method [9], we can achieve a time complexity of $O(2^{n/2})$.

The main benefit of the RFR algorithm is its simplicity, both in theory and in implementation. Our main implementation is a few thousand lines of C code, in comparison to the near 900,000 lines of code for FLINT [10], a contemporary number theory package. This simplicity makes RFR easy to understand and use, and it is also easy to parallelize. Consequently, we have been able to implement the algorithm for massively parallel platforms, such as graphic processing units (GPU), with a 2-3 orders of magnitude improvement in running time.

Our goal in this paper is not to compete with current high-performance algorithms, as this would be a fool's errand! Instead, we want to showcase the power of combinatorial algorithms in the modern era. Starting this project, we set an arbitrary goal of factoring a random polynomial of degree 100. As we will see, the RFR algorithms can factor such a polynomial in a few tenths of a second.

## 2. Mathematical Background

Let $p(x) = x^d + a_{d-1}x^{d-1} + ... + a_1 x + a_0 \in \mathbb{Z}[x]$ be a monic univariate polynomial of degree $d = \deg(p)$. The assumptions of having integer coefficients and being monic are standard. A non-monic polynomial in $\mathbb{Z}[x]$ or $\mathbb{Q}[x]$ can be transformed to this form, factorized, and the factors transformed back to the original form.

In addition, we assume that $p(x)$ is square-free. Therefore, it has distinct roots. Again, this is not a significant restriction. If $p(x)$ is not square-free, we can use Yun's algorithm [11], find a square-free decomposition of $p(x)$, and then apply the factorization algorithm to each square-free factor in turn.

We start by factoring $p(x)$ in $\mathbb{R}$. According to the fundamental theorem of algebra, $p(x)$ has $d$ roots, of which $r$ are real, denoted by $u_i$, and $2c = d - r$ are complex and come in conjugate pairs $z_j$ and $\overline{z_j}$. The factorization of $p(x)$ in $\mathbb{R}$ is

$$p(x) = \prod_{i=1}^{r}(x - u_i)\prod_{j=1}^{c}\left(x^2 - \left(z_j + \overline{z_j}\right)x + z_j\overline{z_j}\right). \tag{1}$$

We should note that in this paper, we assume that we have access to a black-box that gives us the complex roots of a given input polynomial to the desired accuracy. This way, we don't need to be distracted by all the complexities and intricacies of root-finding algorithms.

Let $q(x) \in \mathbb{R}[x] = x^e + b_{e-1}x^{e-1} + ... + b_1 x + b_0$ divides $p(x)$. The set of the roots of $q(x)$ is a subset of the roots of $p(x)$. Therefore,

$$q(x) = \prod_{i \in I}(x - u_i)\prod_{j \in J}\left(x^2 - \left(z_j + \overline{z_j}\right)x + z_j\overline{z_j}\right), \tag{2}$$

where $I$ is a subset of $\{1, ..., r\}$ and $J$ is a subset of $\{1, ..., c\}$.

We aim to find a $q(x)$ so that its coefficients are integer. However, instead of calculating all the coefficients of $q(x)$, we can improve the search speed by focusing on $b_{e-1}$. We have

$$q(x) = x^e - \left(\sum_{i \in I}u_i + \sum_{j \in J}\left(z_j + \overline{z_j}\right)\right)x^{e-1} + .... \tag{3}$$

Therefore,

$$b_{e-1} = -\left(\sum_{i \in I}u_i + \sum_{j \in J}\left(z_j + \overline{z_j}\right)\right). \tag{4}$$

The key to the RFR algorithm is the fact that if a given $q(x)$ is in $\mathbb{Z}[x]$, all of its coefficients, including $b_{e-1}$, are integers.

We call a polynomial $q(x)$ a candidate factor if its roots are a subset of the roots of $p(x)$ and $b_{e-1}$ is an integer. Therefore, to find candidate factors, we define $R = \{u_1, ..., u_r, z_1 + \overline{z_1}, ..., z_c + \overline{z_c}\}$, and then search for subsets of $R$ that sum to an integer.

Note that while $q(x) \in \mathbb{Z}[x]$ implies $b_{e-1} \in \mathbb{Z}$, the converse is not necessarily true. In other words, $b_{e-1} \in \mathbb{Z}$ does not imply $q(x) \in \mathbb{Z}[x]$. To demonstrate this, let us consider Swinnerton-Dyer polynomials:

$$f_n(x) = \prod\left(x \pm \sqrt{2} \pm \sqrt{3} \pm \sqrt{5} \pm ... \pm \sqrt{p_n}\right) \in \mathbb{Z}[x], \tag{5}$$

where $p_n$ is the $n$th prime. Example,

$$f_2(x) = \left(x + \sqrt{2} + \sqrt{3}\right)\left(x + \sqrt{2} - \sqrt{3}\right)\left(x - \sqrt{2} + \sqrt{3}\right)\left(x - \sqrt{2} - \sqrt{3}\right)$$

$$= x^4 - 10x^2 + 1, \tag{6}$$

with roots $\sqrt{2} + \sqrt{3}$, $\sqrt{2} - \sqrt{3}$, $-\sqrt{2} + \sqrt{3}$, and $-\sqrt{2} - \sqrt{3}$. Swinnerton-Dyer polynomials are considered pathological polynomials designed to test factorization algorithms. It is well known that they don't have any proper factor in $\mathbb{Z}[x]$. The RFR algorithm is not immune to them either. Any real factor of a Swinnerton-Dyer polynomial has a dual factor, such that their product has $b_{e-1} = 0$. For example,

$$\left(x - \sqrt{2} - \sqrt{3}\right)\left(x + \sqrt{2} + \sqrt{3}\right) = x^2 - 5 + \sqrt{6}. \tag{7}$$

Therefore, Swinnerton-Dyer polynomials have multiple candidate factors without a true factor.

In summary, the recombination algorithm generates a list of candidate factors $q_1, q_2, ...,$ such that $b_{e-1} \in \mathbb{Z}$ for each $q_i$. Then, the algorithm tests each $q_i$ by calculating all of its coefficients. If $b_j \in \mathbb{Z}$ for $0 \le j \le \deg(q_i)$, $q_i$ is accepted as a true integer factor. This process is similar to the roundabout algorithm, where the presence of a factor in $\mathbb{Z}_m[x]$ is a necessary but not sufficient condition to have the same factor in $\mathbb{Z}[x]$.

In practice, to test whether a candidate factor is an integer polynomial, it is easier to find its traces than calculating all its coefficients. The $i$th trace of a polynomial, denoted as $\mathrm{Tr}_i$, is the sum of $i$th power of its roots [8]. Therefore, $b_{e-1}$ of $q(x)$ is equal to $-\mathrm{Tr}_1(q(x))$. Similar to the argument above, $q(x) \in \mathbb{Z}[x]$ iff all of its traces all also in $\mathbb{Z}$. In this paper, we use this property to test the candidate factors. However, it should be noted that this idea has a central role in the van Hoeij algorithm [8].

Additional simplification is possible. Because we are interested in finding integer subsets, we only need the fractional part of each element in $R$. We define $\boldsymbol{\rho} = (\rho_1, \rho_2, ..., \rho_n)$, where $n = r + c$, $\rho_i = \texttt{frac}\,(u_i)$ for $1 \le i \le r$, and $\rho_{r+j} = \texttt{frac}\,(z_j + \overline{z_j})$, for $1 \le j \le c$. Here, $\texttt{frac} : (\mathbb{R} \to [0, 1))(x) = x \bmod 1 = x - \lfloor x \rfloor$.

Algorithm FACTOR is the top-level factorization algorithm.

1: **function** FACTOR($p$)
2:      ▷ **Input:** $p \in \mathbb{Z}[x]$
3:      ▷ **Output:** $\boldsymbol{\varphi} = \{q_1, q_2, ...\}$ such that $q_i \in \mathbb{Z}[x]$ and $q_i \mid p$
4:      $\boldsymbol{\rho} \leftarrow \emptyset$
5:      **for** $u \leftarrow$ real roots of $p$ **do**
6:          add $\texttt{frac}(u)$ to $\boldsymbol{\rho}$
7:      **for** $z, \overline{z} \leftarrow$ complex roots of $p$ **do**
8:          add $\texttt{frac}(z + \overline{z})$ to $\boldsymbol{\rho}$
9:      $S \leftarrow$ RECOMBINE($\boldsymbol{\rho}$)
10:      $\boldsymbol{\varphi} \leftarrow \emptyset$
11:      **for** $s \leftarrow S$ **do**
12:          $q \leftarrow$ the factor of $p$ corresponding to $s$ ($q \in \mathbb{R}[x]$)
13:          **if** $q \in \mathbb{Z}[x]$ **then**
14:              add $q$ to $\boldsymbol{\varphi}$
15:      **return** $\boldsymbol{\varphi}$

FACTOR calls a recombination algorithm RECOMBINE. This is a placeholder for five algorithms (RECOMBINE_A to RECOMBINE_E, from simpler to faster), which are discussed in the next section.

The input to RECOMBINE is $\boldsymbol{\rho}$. The output is a set of candidate factors, or equivalently subsets of $\boldsymbol{\rho}$, each encoded as an $n$-bit binary number $s \in \{0, 1\}^n = s_n s_{n-1} ... s_2 s_1$. The $i$th bit of $s$ determines whether $\rho_i$ is included in the subset. We call each $s$ a pattern. We define an auxiliary function,

$$\texttt{value} : (\{0, 1\}^n \times [0, 1)^n \rightarrow [0, 1))(s, \boldsymbol{\rho}) = \texttt{frac}\left(\sum_{i=1}^{n} s_i \rho_i\right), \tag{8}$$

which returns the value of a pattern. We test for acceptance of a value as an integer using the following function,

$$\texttt{accept} : ([0, 1) \rightarrow \{0, 1\})(y) = y < \varepsilon \lor (1 - y) < \varepsilon, \tag{9}$$

where $\varepsilon$ is the tolerance parameter. The output of the recombination algorithms is a set $S = \{s_1, s_2, ..., s_i, ...\}$ of patterns, such that $\texttt{accept}(\texttt{value}(s_i, \boldsymbol{\rho}))$ is true.

## 3. Recombination Algorithms

### 3.1. **Algorithm A: Naive Implementation.**

The first algorithm is a naive search implementation. As discussed above, the input is an $n$-element vector $\boldsymbol{\rho} = (\rho_1, ..., \rho_n)$. The output is a set $S$ of candidate patterns. The size of the search space is $2^n$. However, there are two trivial solutions: $s = 0$ corresponding to factor 1 of $p(x)$, and $s = 2^n - 1$, corresponding to $p(x)$ itself.

We can shrink the search space by considering the symmetries of the problem. If $q(x) \in \mathbb{Z}[x]$ is a factor of $p(x)$, so is $p(x)/q(x)$. In other words, $\texttt{accept}(\texttt{value}(s, \boldsymbol{\rho}))$ is true iff $\texttt{accept}(\texttt{value}(\bar{s}, \boldsymbol{\rho}))$ is true, where $\bar{s}$ is the 1-complement of $s$. Therefore, we do not need to search the whole $\approx 2^n$ possibility and can limit the search to 0 to $2^{n-1} - 1$.

```
1:  function RECOMBINE_A(ρ)
2:      ▷ Input: ρ = (ρ₁, ρ₂, ..., ρₙ), where ρᵢ ∈ [0, 1)
3:      ▷ Output: S = {s₁, s₂, ...} such that accept(value(sᵢ, ρ)) is true
4:      n ← length(ρ)
5:      s ← 0
6:      S ← ∅
7:
8:      while s < 2ⁿ⁻¹ do
9:          x ← value(s, ρ)
10:         if accept(x) then
11:             add s to S
12:         s ← s + 1
13:     return S
```

According to the Kac's theorem [12], the expected number of real roots of a random real polynomial of degree $d$ is

$$E_d = \frac{2}{\pi} \ln d + O(1). \tag{10}$$

The length of $\boldsymbol{\rho}$ is $n = r + c$, where $r$ is the number of real roots and $c$ is the number of complex root *pairs*. Therefore,

$$n = r + c \approx E_d + \frac{d - E_d}{2} = \frac{d}{2} + \frac{1}{\pi} \ln d. \tag{11}$$

For large $n$, the average complexity of RECOMBINE_A is $O(2^n) = O(2^{d/2})$. However, if a polynomial has more real roots than expected from Kac's formula, algorithm's performance suffers. The worst case is when all the roots are real. This happens for Swinnerton-Dyer polynomials, which have only real roots.

### 3.2. **Algorithm B: Search-Tree Pruning.**

Algorithm B is a variant of Algorithm A that prunes the implicit search tree and reduces the time complexity. As we will see, we can do much better with Algorithms C-E but in exchange for increasing space complexity. The main point of Algorithm B is to show that the search problem is not as structure-less as it seems.

Algorithm A checks pattern $s + 1$ after checking $s$. In contrast, Algorithm B jumps from $s$ to $s + \Delta s$ if it can prove that none of the patterns in the range $s + 1$ to $s + \Delta s - 1$ can be a solution.

We describe the algorithm using an example. Let $\boldsymbol{\rho} = (0.1, 0.2, 0.4, 0.8, ...)$. Assume the algorithm has just checked $s = s_n s_{n-1}...1000$ and $\texttt{value}(s, \boldsymbol{\rho}) = 0.2$. Based on this value and $\boldsymbol{\rho}$, we can deduce that the next seven patterns cannot be a solution, so $\Delta s = 8$. This is why. Let $s' = s + 7 = s_n s_{n-1}...1111$. We have $\texttt{value}(s', \boldsymbol{\rho}) = \texttt{value}(s, \boldsymbol{\rho}) + \rho_1 + \rho_2 + \rho_3 = 0.2 + 0.7 = 0.9$. Therefore, if $s < x \leq s'$, then $\texttt{value}(s, \boldsymbol{\rho}) < \texttt{value}(x, \boldsymbol{\rho}) \leq \texttt{value}(s', \boldsymbol{\rho})$, or equivalently, $0.2 < \texttt{value}(x, \boldsymbol{\rho}) \leq 0.9$, which means $\texttt{value}(x, \boldsymbol{\rho})$ cannot be an integer. This argument works because chain $\texttt{value}(s, \boldsymbol{\rho}) \to \texttt{value}(s + 1, \boldsymbol{\rho}) \to \to \texttt{value}(s + 7, \boldsymbol{\rho})$ does not cross 1. We can summarize the desired condition as $\texttt{value}(s, \boldsymbol{\rho}) + \rho_1 + \rho_2 + \rho_3 < 1$.

In the example above, to make the jumps as long as possible, $\rho_1$, $\rho_2$, and $\rho_3$ should be as small as possible. Therefore, we assumed that $\boldsymbol{\rho}$ is sorted.

Figure 1 shows a small segment of the search space for a degree 30 polynomial. The x-axis is $s$, and the $y$-axis depicts $\texttt{value}(s, \boldsymbol{\rho})$. The circles mark all the points visited by Algorithm A. On the other hand hand, Algorithm B only visits the closed balls, including the solution (the red ball). Note that the closed balls concentrate near lines 0 and 1 (equivalent because of $\texttt{frac}$) with long jumps in between.

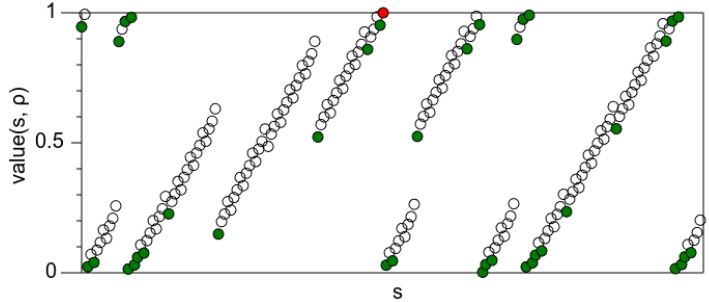

Figure 1. A portion of the search space for a degree 30 polynomial. The x-axis is the pattern ($s$), and the y-axis is the value. Algorithm B visits only the closed circles, unlike Algorithm A, which visits all the circles. The red circle has a value of 1 and is a solution.

recombine_b is presented below. The input is again $\boldsymbol{\rho} = (\rho_1, ..., \rho_n)$. First, $\boldsymbol{\rho}$ is sorted in a non-decreasing order, such that $\rho_1 \leq \rho_2 \leq ... \leq \rho_n$. Let $\boldsymbol{\sigma} = (\sigma_1, ..., \sigma_n)$ be the cumulative sum of $\boldsymbol{\rho}$, defined as $\sigma_0 = 0$ and $\sigma_i = \sigma_{i-1} + \rho_i$ for $i > 0$.

1: **function** RECOMBINE_B($\boldsymbol{\rho}$)
2:     ▷ **Input:** $\boldsymbol{\rho} = (\rho_1, \rho_2, ..., \rho_n)$, where $\rho_i \in [0, 1)$
3:     ▷ **Output:** $S = \{s_1, s_2, ...\}$ such that `accept(value(`$s_i, \boldsymbol{\rho}$`))` is true
4:     sort $\boldsymbol{\rho}$ is a non-decreasing order
5:     $n \leftarrow \text{length}(\boldsymbol{\rho})$
6:     $s \leftarrow 0$
7:     $S \leftarrow \emptyset$
8:     $\boldsymbol{\sigma} \leftarrow$ the cumulative sum of $\boldsymbol{\rho}$
9:
10:     **while** $s < 2^{n-1}$ **do**
11:        $x \leftarrow \text{value}(s, \boldsymbol{\rho})$
12:        **if** `accept(`$x$`)` **then**
13:           add $s$ to $S$
14:        $c \leftarrow$ the count of the trailing zeros of $s$
15:        $j \leftarrow \underset{i}{\arg\max}(x + \sigma_i < 1)$ for $0 \leq i \leq c$
16:        $s \leftarrow s + 2^j$
17:     **return** $S$

We can estimate the average improvement in time complexity expected from Algorithm B.

**Theorem 3.2.1.** *Let $\boldsymbol{\rho} = (\rho_1, ..., \rho_n)$, for $\rho_i \in [0, 1)$. Moreover, $\boldsymbol{\rho}$ is sorted in a non-descending order. The expected ratio of the time complexity of Algorithm A to Algorithm B applied to $\boldsymbol{\rho}$ is $2^l/l$, where $l = \left\lfloor \sqrt{2n} \right\rfloor$.*

*Proof.* We assume that for an integer polynomial with random coefficients, $\rho \in \boldsymbol{\rho}$ is a random variable drawn uniformly from $[0, 1)$. This is because most complex roots are close to the unit circle, combined with the way $\rho$s are generated (especially the effect of `frac`).

For the remaining of this proof, we assume that $\rho$s and $\sigma$s are random variables. Because $\boldsymbol{\rho}$ is sorted, $E(\rho_1) = \frac{1}{2n}$. Similarly, $E(\rho_2) = \frac{3}{2n}$, or more generally, $E(\rho_i) = \frac{2i-1}{2n}$, for $1 \leq i \leq n$. By definition, $\sigma_k = \sum_{i=1}^{k} \rho_i$. Therefore, $E(\sigma_k) = \frac{k^2}{2n}$.

RECOMBINE_B makes jumps of size $1, 2, 2^2, ..., 2^l$, where $\sigma_l < 1$ but $\sigma_{l+1} \geq 1$. From the formula for $E(\sigma_k)$ and setting $E(\sigma_l) = 1$, we have $l = \left\lfloor \sqrt{2n} \right\rfloor$.

Let $m_i$ be the number of jumps of size $2^i$. The **while** loop in RECOMBINE_B runs $m_0 + m_1 + ... + m_l$ time. The improvement ratio we are seeking is

$$\beta = \frac{m_0 + 2m_1 + 4m_2 + ... + m_l 2^l}{m_0 + m_1 + m_2 + ... + m_l}. \tag{12}$$

We need to estimate $m_i$s one by one starting from $m_l$. The algorithm checks $2^l$-wide segments with $l$ trailing zeros. Therefore, the segments are disjointed, with $\frac{2^{n-1}}{2^l}$ of them. Each one is accepted with a probability of $1 - \sigma_l$. Hence,

$$m_l = \left\lfloor \frac{2^{n-1}(1 - \sigma_l)}{2^l} \right\rfloor. \tag{13}$$

Next, we find $m_{l-1}$ using a similar process. However, this time we need to consider the part of the search space already covered by $2^l$ segments.

$$m_{l-1} = \left\lfloor \frac{(2^{n-1} - m_l 2^l)(1 - \sigma_{l-1})}{2^{l-1}} \right\rfloor. \tag{14}$$

Similarly,

$$m_{l-2} = \left\lfloor \frac{(2^{n-1} - m_l 2^l - m_{l-1} 2^{l-1})(1 - \sigma_{l-2})}{2^{l-2}} \right\rfloor. \tag{15}$$

We can continue with this process down to $m_0$ to calculate $\beta$ (Figure 2, the blue curve).

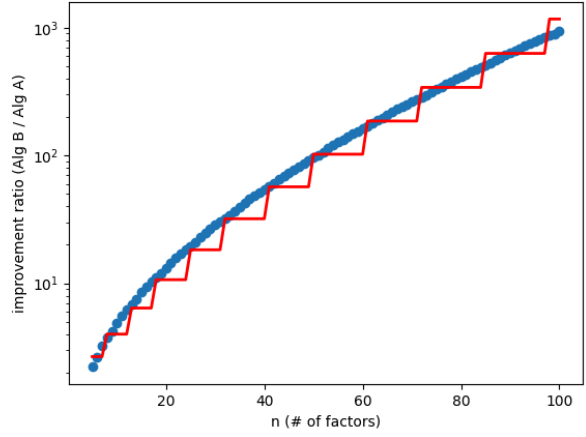

Figure 2. The estimated improvement ratio of Algorithm B compared to Algorithm A. The red curve is the closed-form estimation based on Equation 17.

We can apply a simplification to find an approximate closed formula for $\beta$. In most cases, $m_l$ is relatively small and does not affect $\beta$ significantly. Therefore, we can set it to zero. Moreover, the other $m_i$s are usually of comparable size, say $m$. Therefore,

$$m_0 = m_1 = ... = m_{l-1} = m$$
$$m_l = 0. \tag{16}$$

Substituting in the formula for $\beta$,

$$\beta = \frac{m(1 + 2 + 2^2 + ... + 2^{l-1})}{ml}$$
$$= \frac{2^l - 1}{l} \approx \frac{2^l}{l} \tag{17}$$

The red curve shows the simplified formula.                                    $\square$

### 3.3. **Algorithm C: Space/Time Tradeoff.**

Algorithm B improves on the time-complexity of Algorithm A from $2^n$ to $2^{n-l-\log_2(l)}$, for $l = \lfloor \sqrt{2n} \rfloor$, which is still asymptomatically $O(2^n)$. Now is the time to move to $O(2^{n/2}) = O(2^{d/4})$.

The key idea is that the problem the recombination algorithms are trying to solve is a variant of the *integer subset sum problem*. We have formulated our problem as finding a subset of $\boldsymbol{\rho} = (\rho_i)$, for $\rho_i \in [0, 1)$. Equivalently, we can transform the problem to $\hat{\boldsymbol{\rho}} = (\hat{\rho}_i)$, for $\hat{\rho}_i = \lfloor \rho_i M \rfloor \in Z_M$, $M \in Z > \varepsilon^{-1}$. The auxiliary functions also transform as

$$\texttt{value} : \left(\{0,1\}^n \times Z_M^n \to Z_M\right)(s, \hat{\boldsymbol{\rho}}) = \left(\sum_{i=1}^n s_i \hat{\rho}_i\right) \bmod M, \tag{18}$$

and

$$\texttt{accept} : (Z_M \to \{0,1\})(y) = (y = 0) \vee (y = M - 1). \tag{19}$$

Using this formulation, our problem is to find a subset of $\hat{\boldsymbol{\rho}}$ that sums up to 0 or $M - 1$ using modular addition. We adopt the Horowitz-Sinha algorithm to solve this problem [9]. To stay consistent with the rest of the paper, we keep using real $\boldsymbol{\rho}$ instead of integer $\hat{\boldsymbol{\rho}}$, but the discussion would apply equivalently to both.

The critical insight of the Horowitz-Sinha algorithm is that we can improve the time complexity by a time-space tradeoff. We split $\boldsymbol{\rho}$ into two nearly equal lists: $\boldsymbol{\rho_a} = \left(\rho_1, ..., \rho_{\lfloor \frac{n}{2} \rfloor}\right)$ and $\boldsymbol{\rho_b} = \left(\rho_{\lfloor \frac{n}{2} \rfloor + 1}, ..., \rho_n\right)$. Let $n_a = |\boldsymbol{\rho_a}|$ and $n_b = |\boldsymbol{\rho_b}|$. Of course, $n = n_a + n_b$.

From $\boldsymbol{\rho_a}$, we generate $2^{n_a}$ patterns $s \in \{0, 1, ..., 2^{n_a} - 1\}$ and calculate the corresponding $\texttt{value}(s, \boldsymbol{\rho_a})$. Let $\boldsymbol{\alpha} = (\alpha_1, \alpha_2, ..., \alpha_{2^{n_a}})$ be a list of these values sorted in a non-decreasing order (Algorithm EXPAND). Similarly, we calculate $\boldsymbol{\beta} = (\beta_1, \beta_2, ..., \beta_{2^{n_b}})$ from $\boldsymbol{\rho_b}$.

1:   **function** EXPAND($\boldsymbol{\rho}$)
2:       ▷ **Input:** $\boldsymbol{\rho} = (\rho_1, \rho_2, ..., \rho_n)$, where $\rho_i \in [0, 1)$
3:       ▷ **Output:** $\boldsymbol{\gamma} = (\gamma_1, \gamma_2, ..., \gamma_{2^n})$, where $\gamma_i \in [0, 1)$ and $\gamma_i \le \gamma_{i+1}$
4:       $n \leftarrow \text{length}(\boldsymbol{\rho})$
5:       $s \leftarrow 0$
6:       $\boldsymbol{\gamma} \leftarrow$ array of size $2^n$
7:
8:       **while** $s < 2^n$ **do**
9:           $\gamma_{s+1} \leftarrow \texttt{value}(s, \boldsymbol{\rho})$
10:         $s \leftarrow s + 1$
11:      sort $\boldsymbol{\gamma}$ in a non-decreasing order
12:      **return** $\boldsymbol{\gamma}$

We want to find $\alpha_i \in \boldsymbol{\alpha}$ and $\beta_j \in \boldsymbol{\beta}$ such that $\texttt{accept}(\alpha_i + \beta_j)$ is true. Naively, we can do this by looking all $2^{n_a} 2^{n_b} = 2^n$ different $\alpha_i \beta_j$ combinations. But this is just Algorithm A. Using the Horowitz-Sinha algorithm, we can achieve this in $O(n2^{n/2})$.

The algorithm keeps track of two indices: $i$, $j$. We start by setting $i = 2^{n_a}$ and $j = 1$. Therefore, $\alpha_i$ is the last (largest) element of $\boldsymbol{\alpha}$, and $\beta_j$ is the first (smallest) element of $\boldsymbol{\beta}$. If $\texttt{accept}(\alpha_i + \beta_j)$ is true, meaning $|\alpha_i + \beta_j - 1| < \varepsilon$, we add $(i,j)$ to the list of candidate solutions. Otherwise, if $\alpha_i + \beta_j > 1$, we need to make this sum smaller (closer to 1). We do this by setting $i \leftarrow i - 1$. This works because $\boldsymbol{\alpha}$ is sorted. Conversely, if $\alpha_i + \beta_j < 1$, we set $j \leftarrow j + 1$ to make the sum larger. After the move, we check $\texttt{accept}(\alpha_i + \beta_j)$ again (Algorithm FIND).

1: **function** FIND($\boldsymbol{\alpha}$, $\boldsymbol{\beta}$)
2:     ▷ **Input:** $\boldsymbol{\alpha}$ and $\boldsymbol{\beta}$ are like $\boldsymbol{\gamma}$ in FIND
3:     ▷ **Output:** $S = \{s_1, s_2, ...\}$ such that $s_i$ is an acceptable pattern
4:     $n_\alpha \leftarrow \text{length}(\boldsymbol{\alpha})$
5:     $n_\beta \leftarrow \text{length}(\boldsymbol{\beta})$
6:     $i \leftarrow n_\alpha$
7:     $j \leftarrow 1$
8:     $S \leftarrow \emptyset$
9:
10:     **while** $(i \geq 1) \wedge (j \leq n_\beta)$ **do**
11:         **if** $\texttt{accept}(\alpha_i + \beta_j)$ **then**
12:             $s \leftarrow$ the pattern corresponding to $i$ and $j$
13:             add $s$ to $S$
14:             $i \leftarrow i - 1$
15:         **else if** $\alpha_i + \beta_j > 1$ **then**
16:             $i \leftarrow i - 1$
17:         **else**
18:             $j \leftarrow j + 1$
19:     **return** $S$

Figure 3 shows FIND in action. Finally, we put everything together as RECOMBINE_C.

1: **function** RECOMBINE_C($\boldsymbol{\rho}$)
2:     ▷ **Input:** $\boldsymbol{\rho} = (\rho_1, \rho_2, ..., \rho_n)$, where $\rho_i \in [0,1)$
3:     ▷ **Output:** $S = \{s_1, s_2, ...\}$ such that $\texttt{accept}(\texttt{value}(s_i, \boldsymbol{\rho}))$ is true
4:     $n \leftarrow \text{length}(\boldsymbol{\rho})$
5:     $\boldsymbol{\rho_a} \leftarrow \left(\rho_1, ..., \rho_{\lfloor n/2 \rfloor}\right)$
6:     $\boldsymbol{\rho_b} \leftarrow \left(\rho_{\lfloor n/2 \rfloor + 1}, ..., \rho_n\right)$
7:     $\boldsymbol{\alpha} \leftarrow \text{EXPAND}(\boldsymbol{\rho_a})$
8:     $\boldsymbol{\beta} \leftarrow \text{EXPAND}(\boldsymbol{\rho_b})$
9:     $S \leftarrow \text{FIND}(\boldsymbol{\alpha}, \boldsymbol{\beta})$
10:     **return** $S$

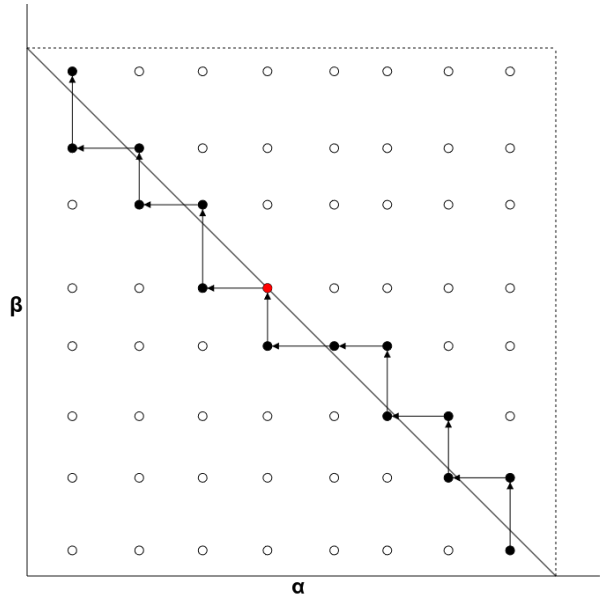

Figure 3. A schematic showing how FIND works. The staircase search pattern stays close to the diagonal line $\alpha + \beta = 1$ until it finds a candidate point on the line (the red circle).

The time complexity of EXPAND is dominant by sorting. Assume that $n_a = n_b = \frac{n}{2}$. Then, the time-complexity of EXPAND is $\frac{n}{2}2^{n/2}$. Because it is called twice, once for $\boldsymbol{\alpha}$ and once for $\boldsymbol{\beta}$, the total time complexity becomes $n2^{n/2}$. The inner loop in FIND runs at most $n_\alpha + n_\beta = 2^{n_a} + 2^{n_b} = 2n^{n/2}$ times. All together, we have $(n + 2)2^{n/2}$ steps or $O(2^{n/2})$. The space complexity is determined by the space needed to store $\boldsymbol{\alpha}$ and $\boldsymbol{\beta}$ and is $O(2^{n/2+1})$.

### 3.4. Algorithm D: Fast Sorting.

The time complexity of Algorithm C is $O(n2^{n/2})$, significantly improving compared to Algorithms A and B. More improvement is still possible. Specifically, we want to remove the leading $n$. This $n$ comes from the sorting algorithm in EXPAND since sorting a $k = 2^n$-element array using a general sorting algorithm is of order $O(k \log(k)) = O(n2^n)$. However, the problem has enough structure, so sorting with a time-complexity linear in the number of elements is feasible.

While discussing Algorithm B, we mentioned that we can assume that $\rho \in \boldsymbol{\rho}$ is a random variable drawn uniformly from $[0, 1)$. For a given input $\boldsymbol{\rho}$, EXPAND finds $k$ subsets of $\boldsymbol{\rho}$, adds the members of the subsets, and then applies `frac`. Therefore, we can assume that the values going into lists $\boldsymbol{\alpha}$ and $\boldsymbol{\beta}$ are also drawn uniformly from $[0, 1)$. This fact allows us to estimate the likely index of the value obtained from a given pattern $s$ in $O(1)$. For example, let $x = $ `value`$(s, \boldsymbol{\rho})$. Then, the index of $x$ in $\boldsymbol{\alpha}$ is $\approx \lfloor kx \rfloor + 1$.

The algorithm starts with an array $\boldsymbol{\gamma}$, where each element is marked as empty ($\boldsymbol{\gamma}$ stands for either $\boldsymbol{\alpha}$ or $\boldsymbol{\beta}$). For now, we assume that the length of $\boldsymbol{\gamma}$ is the same as the corresponding $\boldsymbol{\alpha}$ or $\boldsymbol{\beta}$ in Algorithm C. The algorithm generates all the relevant patterns and, for each pattern, say $s$, it calculates $x = \texttt{value}(s, \boldsymbol{\rho})$. Then, for each $x$, the expected index is estimated as $i = \lfloor kx \rfloor + 1$, where $k$ is the length of $\boldsymbol{\gamma}$. If the $i$th element of $\boldsymbol{\gamma}$ is empty, the algorithm simply copies $x$ to that location. Otherwise, it applies *insertion sort* starting from $i$ to find a place for $x$ (Figure 4, see INSERT below for details). This algorithm is related to bucket and pigeonhole sorting algorithms. In this paper, we call this variant *splat sort*.

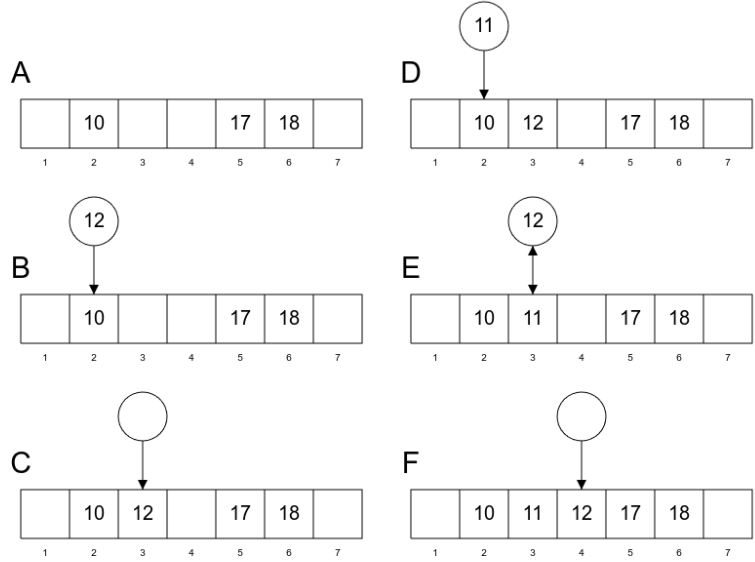

FIGURE 4. Example operations of INSERT. **A** is the starting state. **B** and **C** show how 12 is inserted (collision in **B** and success in **C**). Panels **D-F** depict the insertion of 11. Note that 11 and 12 are exchanged in **E**.

During the early iterations of the algorithm, $\boldsymbol{\gamma}$ is mostly empty. Therefore, it can find an empty spot easily with one attempt. However, as $\boldsymbol{\gamma}$ fills up, finding an empty spot becomes more difficult, and the algorithm must look at multiple spots. Let $l(i)$ be the expected number of spots to look to add the $i$th pattern. Then, the time complexity of the algorithm is $\sum_{i=1}^{k} l(i)$. To calculate $l(i)$, we consider that at this stage, there are already $i-1$ patterns in the array. Therefore, the occupancy ratio is $\mu = \frac{i-1}{k}$. The probability that $x = \texttt{value}(i, \boldsymbol{\rho})$ finds an empty spot on the first attempt is $1 - \mu$. The probability of finding an empty spot after two attempts is $\mu(1-\mu)$. After three spots, we have $\mu^2(1-\mu)$ and so on for the rest. Therefore,

$$\begin{aligned}
l(i) &= (1-\mu) + 2\mu(1-\mu) + 3\mu^2(1-\mu) + ... \\
&= 1 - \mu + 2\mu - 2\mu^2 + 3\mu^2 - 3\mu^3 + ... \\
&= 1 + \mu + \mu^2 + \mu^3 + ... \\
&= \frac{1}{1-\mu}
\end{aligned} \tag{20}$$

Substituting for $\mu$,

$$l(i) = \frac{1}{1-(i-1)/k}. \tag{21}$$

We can estimate the algorithm's time complexity by transforming the sum into an integral,

$$\sum_{i=1}^{k} \frac{1}{1-\frac{i-1}{k}} \approx k \int_{0}^{\frac{k-1}{k}} \frac{1}{1-\mu} \, \mathrm{d}\mu = -k \log\left(1 - \frac{k-1}{k}\right) = k \log(k). \tag{22}$$

Therefore, the time complexity is $O(k \log(k))$, the same as a general sorting algorithm! It seems that we have not made any progress. However, there is a trick to reduce this time complexity to $O(k)$. Most of the algorithm's work is spent on the last few insertions. Therefore, if we sufficiently increase the size of $\boldsymbol{\gamma}$, most insertions will be done with one or two attempts. What is a sufficient size? The following theorem answers this question.

**Theorem 3.4.2.** *The time-complexity of the splat sort for an input of size $k$ and output of size $ck$, for $c > 1$, is $Ck$, where $c \log\left(\frac{c}{c-1}\right) < C < \frac{c}{c-1}$.*

*Proof.* The lower limit is found by a calculation similar to the above argument for when the size of $\boldsymbol{\gamma}$ is $k$. This is a lower limit because the non-empty cells of $\boldsymbol{\gamma}$ are not distributed uniformly and tend to cluster. Therefore,

$$C = \frac{1}{k} \sum_{i=1}^{k} \frac{1}{1-\frac{i-1}{ck}} > \int_{0}^{\frac{k-1}{ck}} \frac{1}{1-\mu} \, \mathrm{d}\mu = c \log\left(\frac{c}{c-1}\right). \tag{23}$$

To find the upper limit, we replace $i$ in the sum with its maximum value. We have,

$$C = \frac{1}{k} \sum_{i=1}^{k} \frac{1}{1-\frac{i-1}{ck}} < \frac{1}{k} \sum_{i=1}^{k} \frac{1}{1-\frac{k-1}{ck}} < \frac{c}{c-1}. \tag{24}$$

$\square$

In this paper, we use $c = 2$. The reason for this choice to simplify the algorithms and allow us to use bit-manipulation tricks instead of multiplication to find indices. For $c = 2$, we can $2 \log(2) < C < \frac{2}{2-1}$; therefore, $1.386 < C < 2$. Experimentally, $C \approx 1.67$ for random polynomials.

The algorithm is summarized below.

1: **function** SPLAT($\boldsymbol{\rho}$)
2:      ▷ **Input:** $\boldsymbol{\rho} = (\rho_1, \rho_2, ..., \rho_n)$, where $\rho_i \in [0, 1)$
3:      ▷ **Output:** $\boldsymbol{\gamma} = (\gamma_1, \gamma_2, ..., \gamma_{2^n})$, where $\gamma_i \in [0, 1)$ and $\gamma_i \leq \gamma_{i+1}$
4:      $n \leftarrow \text{length}(\boldsymbol{\rho})$
5:      $s \leftarrow 0$
6:      $\boldsymbol{\gamma} \leftarrow$ array of size $2^{n+1}$ of empty cells
7:
8:      **while** $s < 2^n$ **do**
9:          $x \leftarrow \texttt{value}(s, \boldsymbol{\rho})$
10:         INSERT($x, \boldsymbol{\gamma}$)
11:         $s \leftarrow s + 1$
12:      compact $\boldsymbol{\gamma}$ by removing the empty cells and truncate $\boldsymbol{\gamma}$ to size $2^n$
13:      **return** $\boldsymbol{\gamma}$

In steps 12, we need to compactify $\boldsymbol{\gamma}$ by removing the empty cells and moving all non-empty cells to the first half of the array. Of course, this can be done easily in linear time. Subroutine INSERT forms the inner loop of the algorithm,

1: **function** INSERT($x, \boldsymbol{\gamma}$)
2:      ▷ **Input:** $x \in [0, 1)$
3:      ▷ **Input:** $\boldsymbol{\gamma} = (\gamma_1, \gamma_2, ..., \gamma_k)$
4:      $k \leftarrow \text{length}(\boldsymbol{\gamma})$
5:      $i \leftarrow \lfloor kx \rfloor + 1$
6:
7:      **while true do**
8:          **if** $\gamma_i$ is empty **then**
9:             $\gamma_i \leftarrow x$
10:            **return**
11:         **if** $\gamma_i > x$ **then**
12:            $\gamma_i \leftrightarrow x$
13:         $i \leftarrow (i \bmod k) + 1$

INSERT uses circular indexing to avoid shifting large blocks of cells, i.e., the index after $i$ is calculated as $(i \bmod k) + 1$. Finally, RECOMBINE_D is essentially the same as RECOMBINE_C, but with SPLAT replacing EXPAND.

1: **function** RECOMBINE_D($\boldsymbol{\rho}$)
2:      ▷ **Input:** $\boldsymbol{\rho} = (\rho_1, \rho_2, ..., \rho_n)$, where $\rho_i \in [0, 1)$
3:      ▷ **Output:** $S = \{s_1, s_2, ...\}$ such that $\texttt{accept}(\texttt{value}(s_i, \boldsymbol{\rho}))$ is true
4:      $n \leftarrow \text{length}(\boldsymbol{\rho})$
5:      $\boldsymbol{\rho_a} \leftarrow \left( \rho_1, ..., \rho_{\lfloor n/2 \rfloor} \right)$
6:      $\boldsymbol{\rho_b} \leftarrow \left( \rho_{\lfloor n/2 \rfloor + 1}, ..., \rho_n \right)$
7:      $\boldsymbol{\alpha} \leftarrow$ SPLAT($\boldsymbol{\rho_a}$)
8:      $\boldsymbol{\beta} \leftarrow$ SPLAT($\boldsymbol{\rho_b}$)
9:      $S \leftarrow$ FIND($\boldsymbol{\alpha}, \boldsymbol{\beta}$)
10:     **return** $S$

Compared to Algorithm C, the time-complexity has improved to $O\left(2^{n/2+1}\right)$, but the space-complexity has slightly worsened to $O\left(2^{n/2+2}\right)$.

### 3.5. Algorithm E: Sort and Query.

Algorithm E aims to improve Algorithm D's space complexity. Along the way, it became simpler and slightly faster.

The critical difference from Algorithm D is that, with some modifications, SPLAT can be used as a query subroutine, i.e., to check whether a given value exists in a sorted array. Therefore, we do not need to generate two separate arrays $\boldsymbol{\alpha}$ and $\boldsymbol{\beta}$. Instead, we can generate the values of the second array on the fly and query into the first one.

The function to test whether a sorted array contains a value is called QUERY.

1:  **function** QUERY($x$, $\boldsymbol{\gamma}$)
2:      ▷ **Input:** $x \in [0, 1)$
3:      ▷ **Input:** $\boldsymbol{\gamma} = (\gamma_1, \gamma_2, ..., \gamma_k)$, where $\gamma_i \in [0, 1)$ and $\gamma_i \leq \gamma_{i+1}$
4:      ▷ **Output: true** if $\exists y \in \boldsymbol{\gamma}$ such that $\texttt{accept}(x - y)$ otherwise **false**
5:      $k \leftarrow \text{length}(\boldsymbol{\gamma})$
6:      $i \leftarrow \lfloor kx \rfloor + 1$
7:
8:      **while true do**
9:          **if** $\gamma_i$ is empty **then**
10:             **return false**
11:         **if** $\texttt{accept}(x - \gamma_i)$ **then**
12:             **return true**
13:         **if** $\gamma_i > x$ **then**
14:             **return false**
15:         $i \leftarrow (i \bmod k) + 1$

QUERY is a modified version of INSERT, and its time complexity is, at worst, equal to the time complexity of INSERT. Therefore, it is bounded by theorem 3.4.2, meaning it is $O(1)$ for an array generated by SPLAT and $c = 2$.

Because QUERY performs the search function, we do not need FIND anymore. As the result, RECOMBINE_E directly calls QUERY. Similar to FIND, we want to find $x \in \boldsymbol{\alpha}$ and $y \in \boldsymbol{\beta}$ such that $\texttt{accept}(x + y)$ is true. The difference is that now $\boldsymbol{\alpha}$ is not explicit. As mentioned before, $\texttt{accept}(x + y)$ means $|x + y - 1| < \varepsilon$, which is why we have QUERY($1 - x, \boldsymbol{\beta}$) in line 13.

```
1:  function RECOMBINE_E(ρ)
2:      ▷ Input: ρ = (ρ₁, ρ₂, ..., ρₙ), where ρᵢ ∈ [0, 1)
3:      ▷ Output: S = {s₁, s₂, ...} such that accept(value(sᵢ, ρ)) is true
4:      n ← length(ρ)
5:      ρₐ ← (ρ₁, ..., ρ⌊n/2⌋)
6:      ρ_b ← (ρ⌊n/2⌋₊₁, ..., ρₙ)
7:      β ← SPLAT(ρ_b)
8:      s ← 0
9:      S ← ∅
10:
11:     while s < 2^⌊n/2⌋ do
12:         x ← value(s, α)
13:         if QUERY(1 − x, β) then
14:             s′ ← the pattern corresponding to x in ρₐ and 1 − x in ρ_b
15:             add s′ to S
16:         s ← s + 1
17:     return S
```

The time complexity of Algorithm E is similar to Algorithm D and is $O(2^{n/2+1})$. The space complexity has slightly improved to $O(2^{n/2+1})$. The main benefit of Algorithm E is that it is well-suited for parallelization, which is the subject of the next section.

## 4. Parallel Implementation

One of the goals of this paper is to develop a factorization algorithm that can be implemented on massively parallel hardware. GPUs are the leading such systems and form the backbone of modern high-performance computing (HPC) and machine learning. While the original purpose of GPUs was as graphic accelerators, they have been used in general computing, especially HPC workloads, for decades, and this is what we focus on here. Our brief description uses the nomenclature specific to Nvidia GPUs, although other GPUs work similarly.

The typical execution model of a GPU is single-instruction, multiple-threads (SIMT) [13]. Each GPU has multiple streaming multi-processors (SP), akin to a traditional CPU. Each SP has hundreds to thousands of compute cores, essentially floating-point arithmetic logic units (ALU). The memory is split between global, shared (among a group of threads), and local (for a single thread). Each SP runs a group of threads using the same code but on different data.

It is possible to achieve a 100x+ performance gain if the code is written with close consideration of how the GPU works. On the other hand, if we simply translate a serial code to a GPU, the gain would be disappointing. A GPU works best for embarrassingly parallel workloads, where threads can work in lockstep but on disjoint pieces of data. To achieve the best result, the threads running together on the same SP (called a warp) must avoid divergent code paths and minimize global memory access.

For example, FIND in Algorithm C is not easily to parallelize. One option would have been to calculate $\boldsymbol{\alpha}$ and $\boldsymbol{\beta}$ on the GPU, transfer them back to the CPU, and run FIND in the CPU. However, memory transfers to and from GPU are one of the main bottlenecks of GPU computing and should be minimized. This is why we introduced QUERY in Algorithm E, which is naturally parallelizable. This way, $\boldsymbol{\beta}$ can stay on the GPU and only the list of candidates ($S$) - which is much shorter than $\boldsymbol{\beta}$ and, therefore, faster to move - must be transferred back to the CPU.

While writing parallel code, one must pay close attention to *race conditions*. INSERT in Algorithm D is prone to race condition. This happens because two or more threads can work on the same index. For example, threads A and B may both reach $\gamma_i$. Thread A finds $\gamma_i$ empty, but before it can write $x$, B executes $\gamma_i \leftarrow x'$. However, A is unaware of this and executes $\gamma_i \leftarrow x$. The result is that $x'$ is overwritten and will be missed from further inclusion in $\boldsymbol{\gamma}$. GPUs have *atomic* instructions to prevent such race conditions. The new version of INSERT is shown below. It is functionally the same as the version in Algorithm D but is modified so that the only memory access is through an atomic exchange operation in line 8.

```
 1: function INSERT(x, γ)
 2:         ▷ Input: x ∈ [0, 1)
 3:         ▷ Input: γ = (γ₁, γ₂, ..., γₖ), where γᵢ ∈ [0, 1) and γᵢ ≤ γᵢ₊₁
 4:     k ← length(γ)
 5:     i ← ⌊kx⌋ + 1
 6:
 7:     while true do
 8:             execute simultaneously y ← γᵢ and γᵢ ← x (atomic exchange)
 9:         if y is empty then
10:                 return
11:         if y > x then
12:                 i ← (i mod k) + 1
13:             x ← y
```

## 5. Results

We compare the performance of the RFR algorithm (specifically Algorithm E) with standard implementations of the Berlekamp-Zassenhaus-van Hoeij algorithm. We have chosen two implementations: sympy (Python) and FLINT (C) [14, 10]. We compare these two with two versions of the RFR algorithm: a single-threaded CPU implementation and a GPU one. Both are written in C/C++/CUDA language. The GPU code runs on a circa 2016 Nvidia GPU (GTX 1080) with 2560 compute cores and 8 GB of memory.

Figure 5 shows the comparison. For an input polynomial of degree $d$, we generate two random monic non-reducible polynomials of degree $\frac{d}{2}$ with coefficients in the range $[-100,100]$ and use their product as the input.

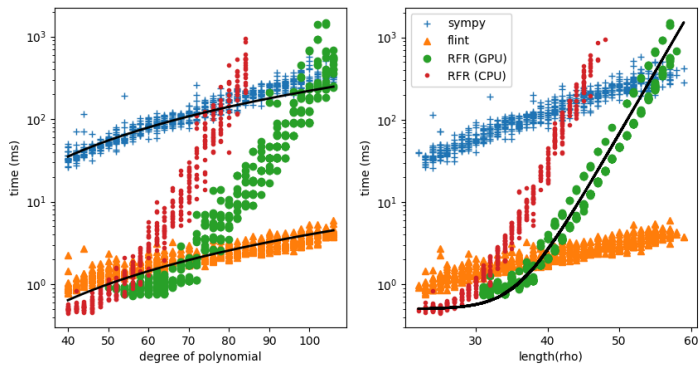

FIGURE 5. Benchmarking the RFR algorithm vs standard factorization implementations.

We have plotted the same data on two different x-axis. On the left, the x-axis is the degree of the input polynomial ($d$). On the right, we use length($\rho$) as the x-axis.

First, compare the sympy (blue crosses) to FLINT (orange triangles) performance. As is well-known and previously reported [15], there is a nearly 100 difference between the two, although they use more or less the same algorithm. The equation of the regression lines on the left panel is $y = 0.022d^2$ for sympy and $y = 4 \times 10^{-4}d^2$ for FLINT, which show the expected quadratic behavior of these algorithms for random polynomials (as mentioned in the Introduction, the Berlekamp-Zassenhaus algorithm is potentially exponential for certain polynomials).

The RFR algorithm shows the expected exponential ($2^{n/2} \approx 2^{d/4}$) time complexity. The regression line on the right is $y = 0.5 + 2 \times 10^{-6}(2^{n/2})$. The GPU implementation (the green circles) is nearly 50x faster than single-threaded CPU implementation (red dots).

Despite the complexity difference, the GPU version of the RFR algorithm is faster than FLINT for $d < 70$ and sympy for $d < 100$. If we use a top-of-the-line GPU (say, GTX4090), which is 10x faster than the one used here, we expect to move the crossing point to $d \approx 120 - 125$.

## 6. Conclusions

In this paper, we introduce a factorization algorithm to factor $p \in \mathbb{Z}[x]$ by recombination of its factors over $\mathbb{R}$. By framing the recombination phase as an integer subset sum problem, we can leverage classic methods, like the Horowitz-Sahni algorithm, to enhance the time-complexity of the algorithm to $O(2^{d/4})$, where $d$ is the degree of $p$.

Despite the algorithm's exponential nature, inherent in the NP complexity group of the integer subset sum, it offers the advantages of simplicity and straightforward parallelization. With the shift in computing towards massively parallel systems, as evidenced by the rise of large AI systems and the trend of the HPC applications, this paper underscores the potential of using massively parallel hardware in solving symbolic algebra problems.

The symbolic algebra algorithms have become more complex and mathematically sophisticated in the last half-century. Polynomial factorization is a case study of this phenomenon, as can be seen from the evolution from Schubert-Kronecker to Berlekamp to Cantor-Zassenhaus to Lenstra-Lenstra-Lovasz and finally to van Hoeij algorithms. However, a disadvantage of such mathematically sophisticated algorithms is the difficulty of implementing them on parallel systems.

This paper argues that sometimes simple and mathematically *less sophisticated* methods can be useful as they can be easily parallelizable. Such simple but parallel algorithms may find application as supplementary and complementary to traditional methods. For example, the RFR algorithm, with its exponential average case, cannot compete with the polynomial time complexity of current factorization algorithms and is unlikely to be a viable standalone factorization algorithm except for small polynomials. However, it is possible that combining it with standard algorithms, for example, by using it as the recombination portion of the roundabout algorithm, may result in a practical and performant hybrid algorithm (we are currently exploring this possibility). A similar strategy may be applicable to other computationally intensive symbolic algebra problems, where part of the algorithm is replaced with an easily parallelizable algorithm that can fully utilize modern computing platforms.

EMORY UNIVERSITY, ATLANTA, GA
*Email address:* shahriar.iravanian@emoryhealthcare.org
*URL:* svtsim.com